\documentclass[%
reprint,
superscriptaddress,
nofootinbib,
nobibnotes,
 amsmath,
 amssymb,
prb,
showpacs,
]{revtex4-1}

\usepackage{graphicx} % Include figure files
\usepackage{dcolumn}  % Align table columns on decimal point
\usepackage{bm}       % bold math

\bibpunct{}{}{,}{s}{}{,}
\begin{document}

\title{Persistence of a Non-Equilibrium State: Observation of a Boltzmannian Special Case\vspace{-1ex}}
\author{D. S. Lobser$^1$}
\author{A. E. S. Barentine$^1$}
\author{E. A. Cornell$^1$}
\author{H. J. Lewandowski\vspace{-5ex}}
\altaffiliation{JILA, National Institute of Standards and Technology and University of Colorado, and Department of Physics, University of Colorado, Boulder, Colorado 80309-0440, USA}
\maketitle
{\bf Well before the atomistic nature of matter was experimentally established, Ludwig Boltzmann's audacious effort to explain the macroscopic world of human experience in terms of the workings of an unseen microscopic world met with vigorous opposition\cite{histnote}.
A contentious point was the problem of irreversibility: the microscopic equations of motion are reversible, yet friction and viscosity cause things always to slow down and warm up, never to speed up and cool down~\cite{Moue2008}.
What was worse, Boltzmann himself discovered that his transport equation predicts special cases in which gases never come to thermal equilibrium, a particular example being that the monopole ``breathe'' mode of gas will never damp if it is confined in 3D to a perfectly isotropic harmonic potential~\cite{BoltzmannOrig}.
Such absences of damping were not observed in nature.
Nondamping of a monopole mode in lower dimensional systems has only very recently been observed, using cold atoms.
Kinoshita et al.~\cite{WeissCradle} and Chevy et al.~\cite{Chevy2002} have experimentally observed suppressed relaxation in highly elongated geometries.
The difficulty in generating sufficiently spherical harmonic confinement for ultracold atoms, however, has meant that Boltzmann's fully 3D, isotropic case has never been observed.
With the development of a new magnetic trap~\cite{LobserThesis} capable of producing near-spherical harmonic confinement for ultracold atoms, we have been able to make the first observation of this historically significant oddity.
We observe a monopole mode for which the collisional contribution to damping vanishes, a long-delayed vindication for Boltzmann's microscopic theory.
}

The Boltzmann equation determines how the phase-space distribution of a gas, $f\left(\mathbf{r},\mathbf{v},t\right)$, evolves as a function of binary collisions between particles with mass, $m$, in the presence of an external force, $\mathbf{F}$
\begin{equation}
\frac{df}{dt} = \frac{\partial f}{\partial t}+\mathbf{v}\cdot\nabla_\mathbf{r}f+\frac{\mathbf{F}}{m}\cdot\nabla_{\mathbf{v}}f =  I_{coll}\left[f\right].
\label{eq:boltzbasic}
\end{equation}
The collision integral, $I_{coll}$, describes how populations at the same location, $\mathbf{r}$, but differing velocities, $\mathbf{v}$, redistribute to two new velocities, $\mathbf{v'}$. 
These local, pairwise collisions conserve momentum and energy.
For the explicit form of the integral, see~\cite{Huang,pottier2010nonequilibrium}.

The collision integral vanishes whenever the product of two single-particle distributions is identical directly before and after a collision, in other words, when 
\begin{equation}
f(\mathbf{v_1})f(\mathbf{v_2}) = f(\mathbf{v_1}')f(\mathbf{v_2}').
\label{eq:equilconstraint}
\end{equation}
This equality typically implies that the phase-space distribution is time-invariant and the gas has reached equilibrium.
However, when energy, momentum, and total atom number are conserved, equation \eqref{eq:equilconstraint} is generically satisfied by distributions of the form
\begin{equation}
f\left(\mathbf{r},\mathbf{v},t\right) \propto U(\mathbf{r})e^{-\frac{|m\mathbf{v} - \eta(\mathbf{r},t)|^2}{2mk_BT(t)}},
\label{eq:noneq}
\end{equation}
where $U(\mathbf{r})$ contains information about the external confining potential, $k_B$ is the Boltzmann constant, the temperature $T(t)$ is time-dependent and $\eta(\mathbf{r},t)$ is an arbitrary function of space and time.
While these distributions, known as ``local equilibrium distributions'', always cause the collision integral to vanish, they in general do not satisfy equation \eqref{eq:boltzbasic}~\cite{pottier2010nonequilibrium,Huang,UhlenbeckFord,Cercignani1988}.
By constraining the local equilibrium distribution so that $\eta(\mathbf{r},t)=0$ and $dT/dt=0$, the distribution becomes a valid solution of equation \eqref{eq:boltzbasic} known as the Maxwell-Boltzmann distribution:
\begin{equation}
f\left(\mathbf{r},\mathbf{v}\right) \propto U(\mathbf{r})e^{-\frac{mv^2}{2k_BT}}.
\label{eq:maxboltz}
\end{equation}
But certain potentials exist where equation \eqref{eq:boltzbasic} is satisfied by \textit{non-equilibrium} distributions, in which case the time-dependence in equation \eqref{eq:noneq} remains~\cite{UhlenbeckFord,Cercignani1988}. 
One of these cases is the 3D isotropic harmonic potential with a solution corresponding to a spherically symmetric ``monopole mode'', where temperature and cloud size oscillate, with opposite phase, in time~\cite{Cercignani1988,Guery-Odelin1999,pitaevskii2003bose-einstein,kohntheorem}
Because the temperature is oscillating in the absence of heat conduction, it is convenient to call the temperature in equation \eqref{eq:noneq} a ``kinetic temperature'', $T_k(t)$, and define a ``spatial temperature'', $T_s(t)$, which determines the variation in cloud size.
The average temperature $(T_k(t) + T_s(t))/2$ is constant; the breathing dynamics are analogous to the oscillatory exchange between kinetic and potential energy that occurs in simple harmonic motion.
While the structure of equation \eqref{eq:boltzbasic} implies a quasicontinuous distribution $f$ and thus very large atom number $N$, we show in the methods section that the result of vanishing damping is preserved as $N$ increases from 1 to 2, and onto an arbitrary meso- or macroscopic number.

This strange absence of damping holds for the monopole mode, but not necessarily for other collective modes.
For the quadrupole mode, a mode in which the radial and axial widths oscillate 180$^\circ$ out of phase, cross-dimensional coupling from collisions causes damping.
In the limit of an interatomic collision rate, $\gamma_{coll}$, that is much smaller than the trap frequency, the quadrupole mode in an isotropic harmonic potential is predicted to damp at a rate~\cite{Guery-Odelin1999}
\begin{equation}
\Gamma_Q\simeq\frac{1}{5}\gamma_{coll}.
\label{eq:quaddamprate}
\end{equation}
We will measure quadrupole damping rates as a baseline for comparison with measured monopole damping rates.

The experiment is performed with $^{87}$Rb atoms evaporatively cooled in a TOP~\cite{Petrich1995,Ensher1998} magnetic trap with harmonic confinement at frequency $\omega = 2\pi (9.03(2) \text{~Hz})$ and equipped with additional magnetic coils that permit the six distinct parameters of a 3D quadratic potential to be adjusted independently~\cite{LobserThesis}. 
We measure dipole sloshing motion of atoms in the trap to determine $\hat{i}$, $\hat{j}$, $\hat{k}$, the principle axes of the confining potential, and their associated trapping frequencies, $\omega_i$, $\omega_j$, $\omega_k$.
We characterize the residual asphericity,  $(\omega_{max}-\omega_{min})/\bar{\omega}$, where $\omega_{max}$, $\omega_{min}$ and $\bar{\omega}$ are respectively the maximum, minimum and mean of $\omega_i$, $\omega_j$, $\omega_k$. 
The residual asphericity drifts with time so we periodically retune and recharacterize the trap to keep asphericity small, typically less than 0.002.
To minimize the undesireable mean-field potential, we work at temperature $T$ well above the Bose-Einstein condensate transition temperature, $T_c$, between $2T_c<T<3T_c$.

\begin{figure}[!h]
    \begin{center}
    \includegraphics[width=76mm]{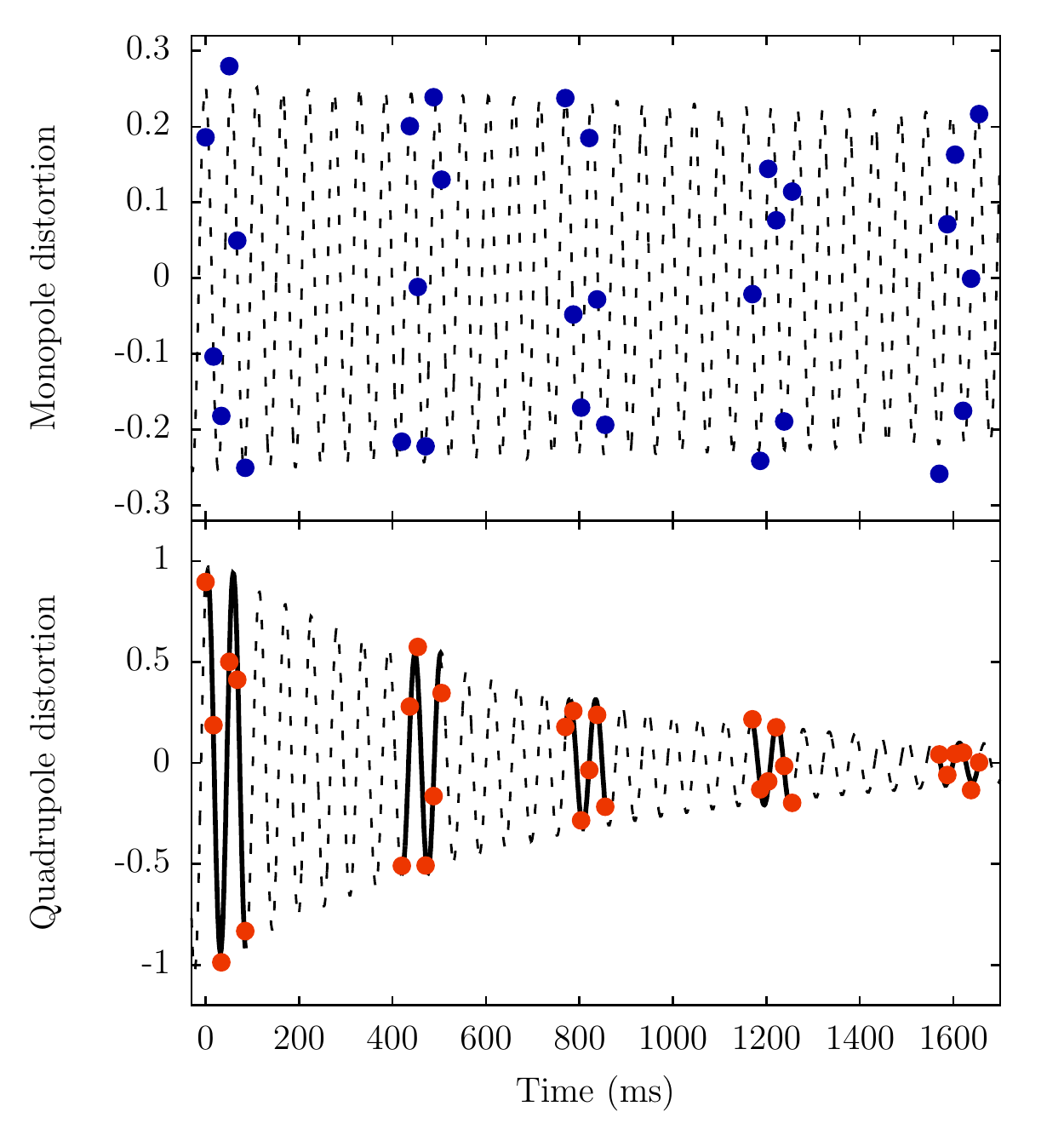}
    \end{center}
    \vspace{-5ex}
    \caption[]{
        Sample data for a driven quadrupole mode and monopole mode in a spherical trap with residual asphericity less than 0.002 and a collision rate of 7.4(3) s$^{-1}$.
        Solid lines on the quadrupole data indicate a typical fitting procedure where individual periods taken in a single run are fit with an undamped sine wave to extract an instantaneous amplitude.
        The oscillation amplitudes at various cycles are in turn fit to an exponential decay to extract the damping rates shown in figure \ref{fig:qbdamp}.
        Only a small subset of the monopole mode data is shown and the full set spans 30 seconds.
        Random observable scatter in these points is predominantly due to small, irreproducible fluctuations in initial conditions.
        \vspace{-5ex}
         }
    \label{fig:quadbreathe}
\end{figure}

We selectively drive monopole (quadrupole) motion by symmetrically (asymmetrically) modulating the strength of the confinement about its mean value.
The cloud is then allowed to evolve freely in the spherical trap before it is non-destructively imaged using phase-contrast microscopy~\cite{haljan,matthews}. 
For each cycle of the experiment, six images are taken of the cloud along two orthogonal axes with an interval of 17 ms in order to sample roughly 1.5 oscillation periods.
Cloud widths along each dimension, $\sigma_{i,j,k}$, are determined using Gaussian surface fits of individual images in order to determine the amplitude of the monopole and quadrupole modes.
Amplitudes of the monopole and quadrupole distortion are scaled by the average width of the cloud during one cycle, given by the following relations
\begin{equation} 
A_M = \frac{\sigma_{i}^2+\sigma_{j}^2+\sigma_{k}^2}{\langle \sigma_i^2+\sigma_j^2+\sigma_k^2\rangle}-1
\end{equation}
\begin{equation}
A_Q = \frac{2\sigma_{k}^2-\sigma_{i}^2-\sigma_{j}^2}{\langle \sigma_i^2+\sigma_j^2+\sigma_k^2\rangle}.
\end{equation}
Oscillation amplitudes are determined by fitting a cycle of oscillation in cloud width, obtained from each experimental run, with a fixed frequency sine wave as indicated by the solid lines in figure \ref{fig:quadbreathe}.

\begin{figure}[!hb]
    \begin{center}
    \includegraphics[width=72mm]{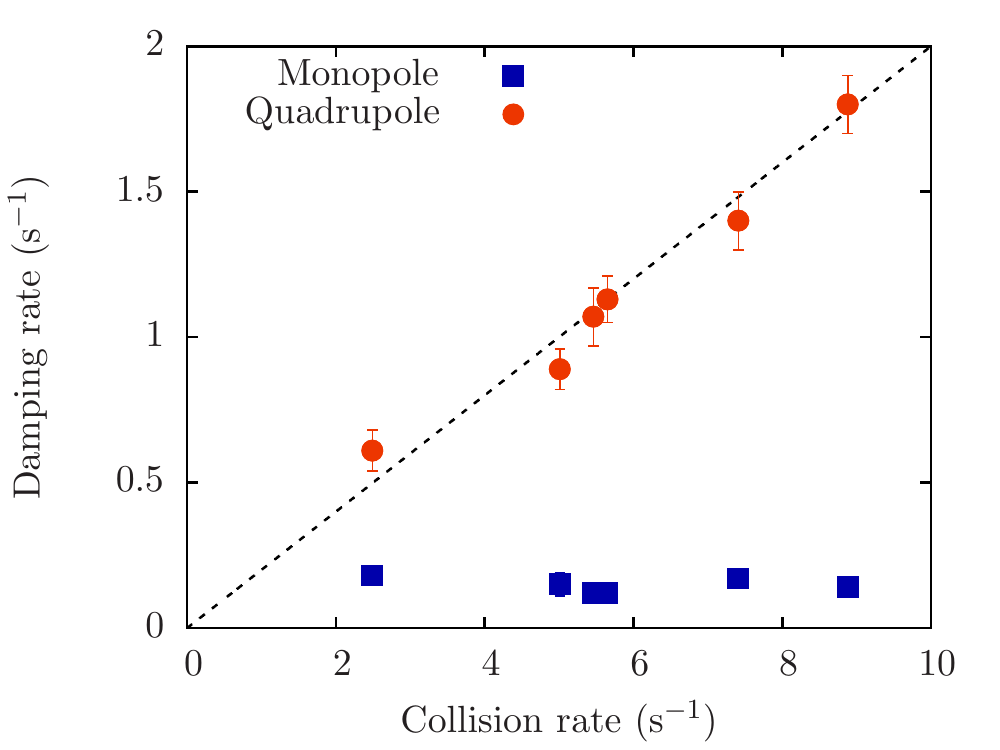}
    \end{center}
    \vspace{-5ex}
    \caption[]{
        Monopole and quadrupole damping rates as a function of interatomic collision rate in a near-spherical trap. 
        The dotted line indicates the predicted quadrupole damping rates given in equation~\eqref{eq:quaddamprate}.
        Residual asphericity was typically around 0.001, but always less than  0.0026. 
        Cloud full width at half maximum was approximately 115 $\mu$m.
        Error bars in the monopole data are smaller than the data points.
        There is a multiplicative uncertainty in the collision rate of 10\%.\vspace{-6ex}
        }
    \label{fig:qbdamp}
\end{figure}

Suppressed damping of the monopole mode relative to the quadrupole mode can be seen in the sample data in figure \ref{fig:quadbreathe}, and although the damping rate is small, it is nonzero.
We characterize the monopole damping by comparing with quadrupole mode damping rates, which are expected to vary linearly with collision rate.
By adjusting the evaporation parameters in our experiment, we can tune $N$, $T$, and the collision rate of the sample, and then alternately drive quadrupole or monopole modes.
A direct comparison of quadrupole and monopole damping rates in a near-spherical trap is shown in figure \ref{fig:qbdamp}. 
The dependence of quadrupole damping on collision rate is $\Gamma_Q = (4.9(1))^{-1} \gamma_{coll}$, which is in good agreement with equation \eqref{eq:quaddamprate}.
The small amount of residual damping in the monopole mode is \textit{independent} of collision rate and much smaller than the damping in the quadrupole mode, as expected.
This, then, is the special-case exception that proves the general rule of damping in the Boltzmann equation.\vspace{-3ex}

In order to understand the source of residual monopole damping, we note that Boltzmann's result hinges on the assumption that the potential is both isotropic and harmonic.
An actual physical system can never satisfy both of these conditions perfectly, and the remainder of this letter is devoted to a discussion of the effects that small anisotropies and anharmonicities have on the monopole damping rate.

Certain subtleties arise for gases in the collisionless limit when anisotropies in the potential are small enough that trap frequencies differ by less than a few percent.
In a \textit{totally} collisionless system, oscillations along the principal axes of the trap are fully decoupled and monopole- or quadrupole-like oscillations are undamped.
If the principal trap frequencies differ such that $\omega_i=\omega_j\neq\omega_k$, dephasing occurs between oscillations along different principal axes and energy exchange between pure monopole motion and pure quadrupole motion occurs with a period given by
\begin{equation}
T_{MQ} = \frac{\pi}{\vert\omega_{k}-\omega_{i}\vert}.
\end{equation}
\noindent
When collisions are included, the two modes become coupled and, as the population in the quadrupole mode increases, so does the damping~\cite{Buggle2005}.
This effect can be seen when $T_{MQ} < 1/\Gamma_Q$, where multiple oscillations between monopole and quadrupole modes occur.
\begin{figure}[!hb]
    \begin{center}
    \vspace{-3ex}
    \includegraphics[width=72mm]{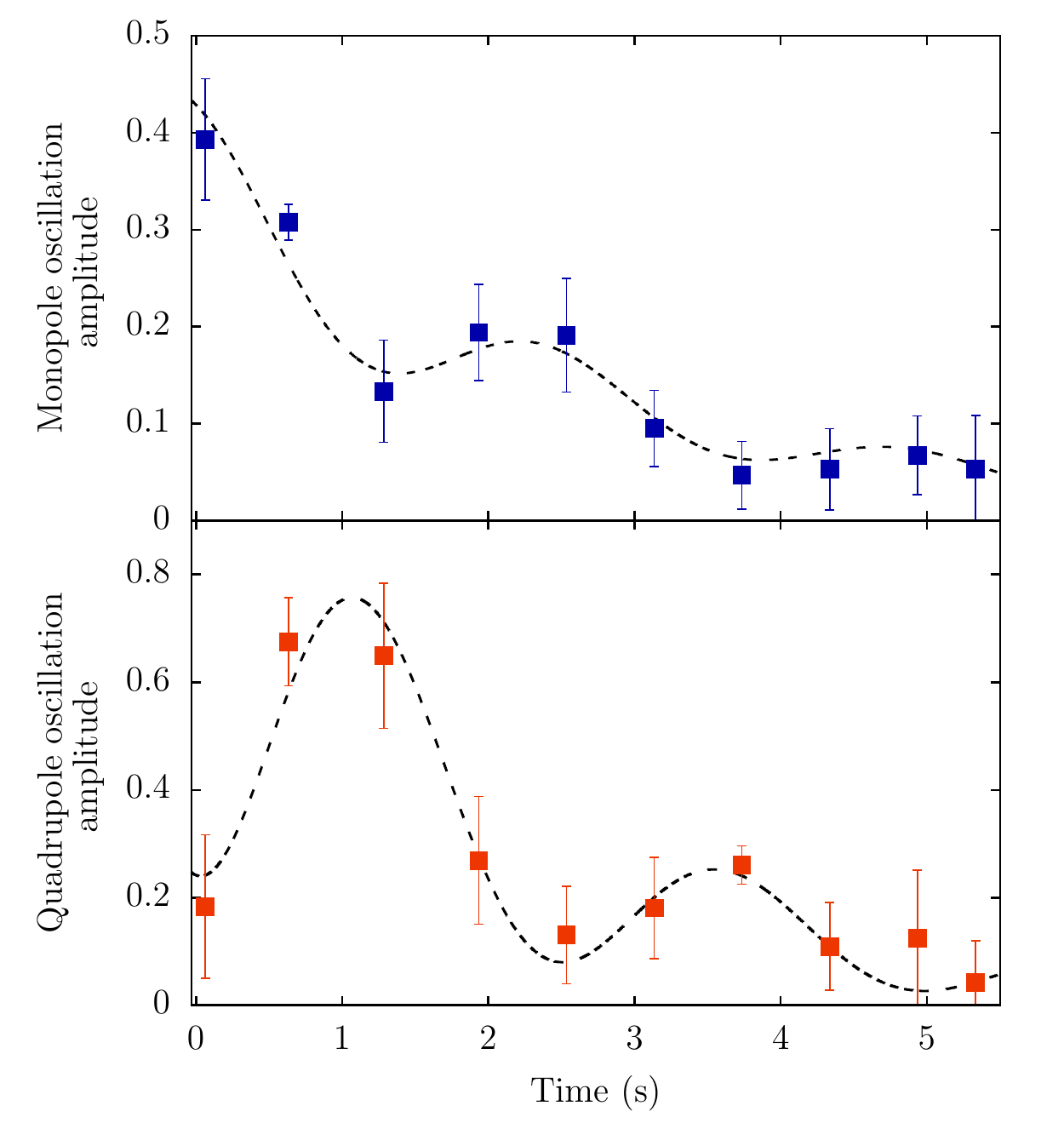}
    \vspace{-4ex}
    \end{center}
    \caption{
       Monopole and quadrupole oscillation amplitudes in an anisotropic trap where the residual asphericity is approximately 0.02.
       The data points show the amplitudes of individual oscillations of the instantaneous distortion, such as those indicated by the black lines in Fig.~\ref{fig:quadbreathe}.
       A monopole mode is initially driven and oscillations between the monopole and quadrupole modes at a frequency $\Delta\omega$ can be seen. 
       The dotted line fitting functions are $e^{-\Gamma t}\left(\cos^2(\Delta\omega t+\delta) + C\right)$.
       Amplitudes are determined from sine-wave fits to individual oscillation measurements obtained from a single experimental cycle. 
       The error bars are calculated from scatter in the sine-wave fits of data taken at repeated evolution times. %\vspace{-4ex}
    }
    \label{fig:qbbeat}
\end{figure}
Data in figure \ref{fig:qbbeat} show oscillations between monopole and quadrupole modes resulting from an initial monopole drive in a trap with residual asphericity of approximately 0.02.
Energy transfers back and forth between the individual modes and the damping rates for both modes are nearly equal, with a mean value of $\Gamma=0.36(4)$ s$^{-1}$.
The collision rate is roughly 3.7 s$^{-1}$ leading to an expected quadrupole damping rate of 0.74 s$^{-1}$ in a spherical trap, which is twice the value of the measured damping rate in the anisotropic trap.
This is no surprise because the quadrupole mode is effectively populated only half of the time, leading to the factor of 2 decrease in the damping rate. 
If we decrease the amount of anisotropy such that $T_{MQ}>>1/\Gamma_Q$, the energy in the quadrupole mode damps before it can fully couple back into the monopole mode.
One can see this effect in the data for the very spherical case shown in figure \ref{fig:quadbreathe}, where the quadrupole mode damps before it can exchange with the monopole mode.
The data in figure \ref{fig:qbdamp} were taken in traps with residual asphericities ranging from 0.0005--0.0026, corresponding to $10.7 \text{ s} < T_{MQ}/2 < 55.6 \text{ s}$. 
The typical relaxation time for the monopole mode in these traps is $\tau_M = 7(2)$ s with no systematic dependence on the residual anisotropy observed.
Thus, some other physical effect must provide the dominant source of residual monopole damping.

We now come to the second condition of Boltzmann's result, which requires that the potential be harmonic, and discuss the effect of anharmonic perturbations to our trapping potential as a source of damping. 
Amplitude-dependent frequency shifts caused by anharmonic perturbations lead to dephasing of individual particle trajectories, effectively damping the collective monopole amplitude.
\begin{figure}[!ht] 
    \begin{center}
    \vspace{-3ex}
    \includegraphics[width=76mm]{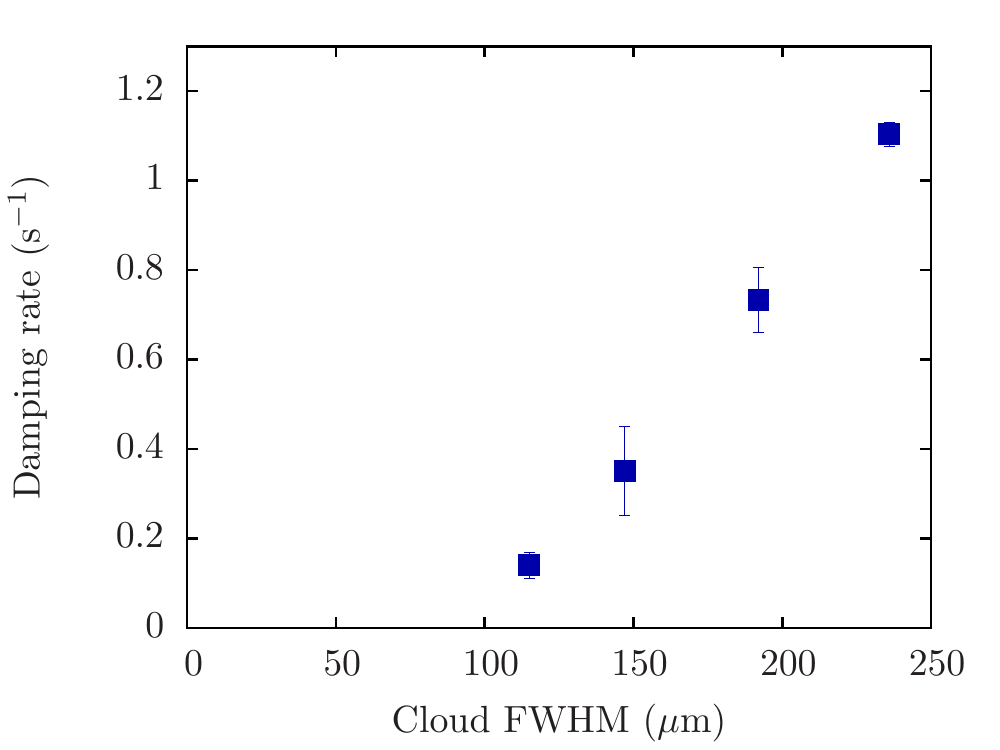}\vspace{-5ex}
    \end{center}
    \caption[]{
        Damping of the monopole mode in a near-spherical trap as a function of the spatial extent of the atom cloud.  The fractional amplitude of the excitation is the same for all points.  Consistent with anharmonicity-induced damping, we see the observed damping rate decreases rapidly for smaller clouds, but the trend of the data suggest that even for the smallest clouds (such as those used in figure \ref{fig:qbdamp}) the residual damping observed may be due to the onset of anharmonic effects.
        The error bars are calculated using the same method described in Fig.~\ref{fig:qbbeat} and the error in the cloud width measurements were typically 2\%.\vspace{-3ex}
    }
    \label{fig:anharm}
\end{figure}
Moreover, the anharmonic corrections to our potential are asymmetric, giving rise to an amplitude-dependent anisotropy. 
A calculation of the expected damping rate that takes into account all of the relevant anharmonic corrections is difficult.
But the effect can be explored experimentally by measuring the monopole damping as a function of cloud size as shown in figure \ref{fig:anharm}.
The first point in figure \ref{fig:anharm} represents the average of the cloud size and monopole damping data in figure \ref{fig:qbdamp}.
The trend of the data in figure \ref{fig:anharm} suggests that the residual damping seen even at the smallest cloud size of 115 $\mu$m is already due to the onset of anharmonic effects.
Unfortunately, we are unable to work with smaller, and thus colder, clouds due to limitations in the signal-to-noise ratio of our imaging system, and the need to keep $T \gtrsim 2T_c$.

In this paper, we present an experimental verfication of the absence of damping for the monopole mode of a thermal gas in an isotropic harmonic potential.
While the damping is highly suppressed, the small but finite relaxation of the monopole mode is an artifact of small anharmonic perturbations to our trap, which decrease with cloud size.
We find that, in the limit of zero anharmonic shifts, the damping of the monopole mode vanishes, as predicted by Boltzmann in 1876~\cite{BoltzmannOrig}.

\section*{methods}
The undamped nature of the monopole mode is found by calculating the evolution of the square radius of the cloud and can be derived in various ways~\cite{Guery-Odelin1999}.
It is instructive to see how the monopole nondamping result can be built-up starting from $N = 1$, $2$, etc\ldots~ 
In spherical symmetry, the radial motion of a single particle of mass $m$, energy $E$, and angular momentum $L$ is governed by the effective potential~\cite{thornton2004classical}
\begin{equation}
V_e = \frac{L^2}{2mr^2}+\frac{1}{2}m\omega^2r^2
\label{eq:effpot}
\end{equation}
so that the radial force is
\begin{equation}
m\frac{d^2r}{dt^2} = -\frac{d}{dr}V_e
\label{eq:radialforce}
\end{equation}
and the kinetic energy is
\begin{equation}
\frac{1}{2}m\left(\frac{dr}{dt}\right)^2 = E-V_e.
\label{eq:kinnrg}
\end{equation}
We note that $d^2r^2/dt^2 = 2(dr/dt)^2+2rd^2r/dt^2$, and substituting \eqref{eq:radialforce} and \eqref{eq:kinnrg} yields the differential equation for $r^2$
\begin{equation}
\label{eq:brth2prt}
\frac{d^2}{dt^2}r^2 = -\Omega^2\left(r^2-r_0^2\right),
\end{equation}
where $\Omega\equiv2\omega$ and $r_0^2=E/(m\omega^2)$. So the square radius undergoes sinusoidal oscillations, or ``monopole breathe'', around its mean value $r_0^2$ at a frequency of $2\omega$. If there are two particles, 1 and 2, each with individual values of $E$, $L$, and $r^2$, each particle will oscillate at $2\omega$.
Taking the sum of their respective differential equations \eqref{eq:brth2prt} yields
\begin{equation} 
\label{eq:monopolecomb}
\frac{d^2}{dt^2}r_t^2 = -\Omega^2\left(r_t^2-r_{0t}^2\right)
\vspace{.5ex}
\end{equation}
where their combined square radius, $r_t^2 \equiv r_1^2+r_2^2$, oscillates around its mean value, $r_{0t}^2 \equiv (E_1+E_2)/(m\omega^2)$.
The magnitude of the collective breathe motion depends on the magnitude and relative phase of the individual particle trajectories.
These individual quantities will abruptly change in the event of a collision.
Assuming the collisions are local, $r_1$, $r_2$, and thus $r_t^2$ will not change from the instant before to the instant after the collision.
Similarly, momentum and energy conservation imply that $\frac{d}{dt}r_t^2$ and $r_{0t}^2$ are unchanged by the collision.

These three continuities imply that the parameters and boundary conditions of equation \eqref{eq:monopolecomb} are matched directly before and after a collision.
This ensures that neither the magnitude nor phase of the oscillation will change as the result of a pairwise collision.
If we instead consider N atoms where
\begin{subequations}
\label{eq:Natoms}
\begin{equation}
r_t^2 = \sum_{i=1}^Nr_i^2
\end{equation}
\begin{equation}
r_{0t}^2 = \frac{1}{m\omega^2}\sum_{i=1}^NE_i,
\end{equation}
\end{subequations}
one can see that the monopole mode is left unperturbed---and in particular \textit{undamped}---by local, pairwise, momentum-, energy-, and number-conserving collisions.
This argument is robust to quantum statistics---Bose or Fermi---and, interestingly, in Ref.~\citenum{Guery-Odelin2014}, it is shown that a $1/r^2$ term in the potential also preserves monopole motion.

The above approach is consistent with a system in the hydrodynamic limit, with total number of atoms so large that the function $f(\mathbf{r},\mathbf{v},t)$ is essentially continuous.
However, in the hydrodynamic limit, mean-field effects can come into play in which case the monopole frequency is shifted~\cite{Guery-Odelin2002}.
In our experiment, the total number of atoms, $N$, is only a few hundred thousand, and the mean-free path is large compared to the spatial extent of the sample---we are not in the hydrodynamic limit in any sense of the word---and mean-field effects can be neglected.

\subsection*{Acknowledgements}
This research was supported by the Marsico fund and NSF grant PHY-1125844.
\bibliographystyle{unsrt}

%\bibliography{ThermalBreathePaperNatureArXiv}

\begin{thebibliography}{10}
\bibitem{kohntheorem}
It is worth noting that the dipole mode is also undamped in a harmonic trap as
  a consequence of Kohn's theorem. However, this result doesn't require that
  the potential be isotropic.
\end{thebibliography}

\end{document}